# Topological-Metal to Band-Insulator Transition in (Bi$_{1-x}$In$_x$)$_2$Se$_3$ Thin Films

Matthew Brahlek[1], Namrata Bansal[2], Nikesh Koirala[1], Su-Yang Xu[3], Madhab Neupane[3], Chang Liu[3], M. Zahid Hasan[3] and Seongshik Oh[1,*]

[1]Department of Physics & Astronomy, Rutgers, The State University of New Jersey, Piscataway, New Jersey 08854, U.S.A.

[2]Department of Electrical and Computer Engineering, Rutgers, The State University of New Jersey, Piscataway, New Jersey 08854, U.S.A.

[3]Joseph Henry Laboratories of Physics, Princeton University, Princeton, NJ, 08544, U.S.A.

*Correspondence should be addressed to ohsean@physics.rutgers.edu

**Abstract:** **By combining transport and photo emission measurements on (Bi$_{1-x}$In$_x$)$_2$Se$_3$ thin films, we report that this system transforms from a topologically non-trivial metal into a topologically trivial band insulator through three quantum phase transitions. At x ≈ 3-7%, there is a transition from a topologically non-trivial metal to a trivial metal. At x ≈ 15%, the metal becomes a variable-range-hopping insulator. Finally, above x ≈ 25%, the system becomes a true band insulator with its resistance immeasurably large even at room temperature. This material provides a new venue to investigate topologically tunable physics and devices with seamless gating/tunneling insulators.**

A new paradigm for classifying condensed matter systems by topology has spurred interest in a new class of materials called topological insulators (TI) [1-3]. At the interface between two insulators that are described by different topological invariants, the energy states are required to become gapless, leading to topologically-protected metallic surface states (SS). For example, $Bi_2Se_3$, a notable topological insulator, exhibits spin-momentum-locked Dirac-like surface bands [4-8] as a result of band inversion induced by strong spin orbit coupling (SOC) [8]. Evidence of SS in TIs has been first observed with surface sensitive probes such as angle resolved photo-emission spectroscopy (ARPES) [9-12] and scanning tunneling microscopy [13], and more recently with transport measurements [14-25].

Here we report transport and ARPES measurements on thin films of $(Bi_{1-x}In_x)_2Se_3$, where x varies from 0 to 1. $Bi_2Se_3$ (x=0) is a proto-typical topological insulator. However, its bulk Fermi energy ($E_F$) always stays in the conduction band (CB), and thus it is better to call $Bi_2Se_3$ a topological metal (TM), as we do in this Letter from now on. On the other hand, $In_2Se_3$ (x=1) is a topologically trivial band insulator (BI) with energy gap around 1.3 eV [26]. Although metal-insulator transitions occur in many other systems [27], and a topological metal to non-topological metal transition has been observed in $TlBi(Se_{1-x}S_x)_2$ [28-29], so far no material system has exhibited tunability from a topological metal state to a truly insulating state. Here we show that $(Bi_{1-x}In_x)_2Se_3$ is the first system to show such a broad tunability.

For a TM to transform into a BI, there have to be at least two quantum critical points. As Bi atoms are replaced by lighter In atoms and the SOC weakens, a quantum critical point will emerge where the energy bands invert, and the system will go from a topologically non-trivial state to a topologically trivial state. At another quantum critical point, $E_F$ will drop below the CB minimum and the system will become an insulator. Depending on which one occurs first, a TM could become a non-topological normal metal (NM) and then a BI, or it could become a TI first and then a BI. We find that the former is the case in this material. In addition to these two quantum phase transitions, we also find that the material goes through a variable range hopping insulator (VI) phase before becoming a BI.

We prepared thin films of $(Bi_{1-x}In_x)_2Se_3$ by molecular beam epitaxy (MBE). The Bi and In source fluxes were calibrated in-situ with a quartz crystal micro-balance, and ex-situ with Rutherford back scattering (RBS) measurements; together these provide accuracy to within about $\Delta x \approx \pm 1\%$. The thickness of the films was fixed at 60 QL (QL, where 1 QL $\approx$ 1 nm), and were grown on $10 \times 10$ mm$^2$ $Al_2O_3$ (0001) substrates (see ref. [30]). Transport measurements were carried out down to 1.5 K in a magnetic field up to 9 T using the standard Van der Pauw geometry. The ARPES measurements were performed at the Advance Light Source in the Lawrence Berkeley National Laboratory. Both ARPES and transport samples were grown in identical conditions except the ARPES samples were capped in-situ by ~100 nm-thick Se. The Se capping was removed in the ARPES chambers by heating the samples up to ~250 °C while keeping the chamber pressure below $1 \times 10^{-9}$ Torr. ARPES measurements were performed at 20 K with 29-64 eV photon energy, with the chamber pressure less than $5 \times 10^{-11}$ Torr.

Figure 1 shows how the structure of $(Bi_{1-x}In_x)_2Se_3$ changes with x. The reflection high energy electron diffraction (RHEED) images in Fig. 1(a) and the X-ray diffraction data in Fig. 1(c) show that all our samples are of a single phase within the instrument limit. In Fig. 1(b), the in-plane lattice constants estimated from the RHEED images gradually decreased from 4.14 Å for $Bi_2Se_3$ to 3.95 Å for $In_2Se_3$.

R(T) curves in Fig. 2(a) show how the system changes from a metallic state to an insulating state as x increases. Samples with x < 15% show metallic temperature dependence, whereas for x $\gtrsim$ 15%, the R(T) curves turn upward as temperature decreases. To see this, we plotted R(1.5 K)/R$_{min}$ vs. x in Fig. 2(b); for x > 18%, R$_{min}$= R(270K) because R(T) monotonically decreased with increasing temperature below 270 K, the highest temperature used in this experiment. Fig. 2(b) shows that R(1.5 K)/R$_{min}$ remains unity for x ≲ 15% and it sharply increases for larger x, which indicates that a metal-to-insulator transition occurs at x = 15%: we define this composition as $x_I$. Although this analysis suggests that the system becomes an insulator for x > 15%, ARPES in Fig. 2(f) shows that the Fermi level remains in the conduction band up to x > 20%.

Such an insulating state is best described by the variable range hopping (VRH) mechanism [31]. With disorder, even if a band is partially filled, electrons can be localized as temperature decreases. According to the VRH model, R(T) is described by $R(T) = R_0\exp[(T_0/T)^{1/(d+1)}]$ where $T_0$ is a characteristic temperature scale for hopping, d is the effective dimensionality of the conducting channel, and $R_0$ is the characteristic resistance of the sample. For x > 15%, the mean free path (Fig. 3(b)) is less than ~1 nm, much smaller than the film thickness (60 nm). In this case, the resistance is dominated by the bulk scattering effect with negligible surface contribution, and the film behaves effectively as a 3d system. Figure 2(d) indeed shows that the temperature dependence of the resistance follows this 3d-VRH formula quite nicely for x > 15%.

From the fitting parameter $T_0$, plotted in Fig. 3(e), we can further check the validity of the 3d-VRH model as a function of x. Because $T_0 \propto 1/N(E_F)$, where $N(E_F)$ is the density of states at the Fermi level [31], as the Fermi level approaches the bottom of the CB and $N(E_F)$ reduces to zero with increasing x, $T_0$ diverges and eventually the VRH mechanism breaks down. Once the Fermi level drops below the CB minimum, the system becomes a BI and the temperature dependence of the resistance should switch to the activation formula, $R(T) = R_0\exp(E_a/k_BT)$, where $k_B$ is Boltzmann's constant, and $E_a$ is the energy difference between the CB minimum and $E_F$. Considering that the band gap of $Bi_2Se_3$ ($In_2Se_3$) is ~0.3 eV (~1.3 eV), any energy scale governing their transport properties cannot be larger than ~1 eV (or ~10,000 K). The 3d-VRH fitting provides $T_0$ =500, 5000, and $6.5 \times 10^5$ K for x = 20, 25 and 30%. This implies that the VRH mechanism is valid for x = 20%, but it is on the verge of breaking down for x = 25% and clearly invalid for x = 30%. On the other hand, $E_a/k_B$ obtained from the activation fitting for x = 30% is 300 K (≈ 30 meV), and this is physically valid considering the band gap shown in Fig. 3(f). Fitting the low temperature data of x = 25% (Fig 2(c)) to the activation formula yields $E_a/k_B \approx 10$ K ($\approx 1$ meV). Considering that this activation energy is larger than the minimum measured temperature (1.5 K), the activation fitting does not violate self-consistency for the 25% data. In other words, x = 25% can barely belong to either side of the VI or the BI within our measurement limit. Summarizing these transport and ARPES analyses, we conclude that x = 20% is a VI, x = 25% is on the boundary between a VI and a BI and x = 30% is a BI.

For x > 30% the samples are immeasurably insulating even at room temperature. The metal-insulator transition in these samples is also observable in their optical transparencies: in Fig. 2(f), 60 QL $Bi_2Se_3$ is completely opaque and mirror-like, whereas $(Bi_{0.2}In_{0.8})_2Se_3$ of the same thickness is semi-transparent even if the band gap of $In_2Se_3$ is smaller than the energy of visible light.

Figure 3(a) presents the results obtained from the Hall effect measurements. Below x = 6%, the Hall resistance, $R_{xy}$, versus B was nonlinear, which indicates the presence of multiple channels contributing to the Hall effect (see our previous paper [20]). This non-linearity disappears above 6% as shown in the inset of Fig. 3(a), where $dR^N_{xy}/dB$ ($R^N_{xy} = R_{xy}(B)/R(9\ T)$) is used as a measure of the non-linearity for x = 2, 6, and 8%. The disappearance of the non-linearity occurs near a point where the surface band disappears in the ARPES spectra (Fig. 3(f)), and is likely a signature of the topological phase transition. Extracting individual carrier densities from non-linear Hall effect data requires a multi-carrier fitting process. To avoid such a complication, we took the slope of the $R_{xy}$ vs. B curves at high field, which is inversely proportional to the sum of the sheet carrier densities, to obtain the total sheet carrier density, $n_{2d}$, (right axis) and the mobility, μ, (left axis) at 1.5 K. In Fig. 3(a), the carrier density monotonically decreased with increasing x, and became too small to be properly measured for x > 20%; thus $n_{2d}$ for 25 and 30% were simply taken to be zero. One notable feature in Fig. 3(a) is that the mobility drops fast right above x = 6% as the non-linearity disappears in $R_{xy}(B)$. This accelerated mobility drop is possibly due to enhanced backscattering as the topological surface bands disappear.

Several key factors must be considered to understand this topological phase transition. In the case of TlBi(Se$_{1-x}$S$_x$)$_2$ [28-29], the replacement of Se by S weakened the SOC and contracted the lattice constant, which resulted in the band inversion and the topological transition at x ≈ 50%, while the bonding orbital (p) characters remain unchanged with x. On the other hand, (Bi$_{1-x}$In$_x$)$_2$Se$_3$ system has, in addition to the SOC and the lattice constant effects, the bonding orbitals change with x: Bi (6p$^3$) has only p orbitals involved in the bonding, whereas In (5s$^2$5p$^1$) has both s and p orbitals. This orbital mixing may have a significant effect on the topological transition of (Bi$_{1-x}$In$_x$)$_2$Se$_3$, considering that the transition occurs at a much smaller value of x than expected.

Additionally, the effect of finite sample thickness has to be considered. In an ideal TI, the penetration depth of the SS into the bulk is given by $\lambda_{ss} = \hbar v_F/E_g$, where $v_F$ is the Fermi velocity of the Dirac band and $E_g$ is the bulk band gap [32]. As x approaches the topological transition point ($x_T$), $E_g$ becomes smaller and $\lambda_{ss}$ increases. If $\lambda_{ss}$ approaches the film thickness, the wavefunctions of the top and bottom surface states should start overlapping with each other, leading to a gap opening at the Dirac point even before the bulk band gap closes and surface band completely disappears [33]. This expectation is indeed consistent with our observation: a gap opened in the ARPES spectrum at x ≈ 3% (Fig. 3(f)), whereas the signature of the surface band at the Fermi level survived up to x ≈ 7% in the transport properties (Fig. 3(a)). In this case, $x_T$ is not a single point, but rather should be defined broadly over ~3-7%, whose sharpness may also depend on the sample thickness. In order to fully confirm this scenario, detailed thickness-dependent ARPES studies will be needed.

The mobility (μ) and carrier density ($n_{2d}$) obtained from the Hall effect allows calculation of the mean free path, $l = (\hbar\mu/e)(3\pi^2 n_{2d}/t)^{1/3}$, where t is the film thickness; l decreases monotonically with x as shown in Fig. 3(b). According to the Ioffe-Regel criterion [34], a material remains metallic only if $k_F l \geq 1$, where $k_F = (3\pi^2 n_{2d}/t)^{1/3}$ is the 3d Fermi wave vector. Fig. 3(b) shows that $k_F l$ continuously diminishes as x increases, and crosses the Ioffe-Regal criterion of $k_F l = 1$ exactly at the composition (x = 15%) where the metal-insulator transition occurs in Fig. 2(b).

Figure 3(d) shows two separately estimated Fermi levels ($E_F$) relative to the bottom of the CB from transport and ARPES measurements. The transport $E_F$ is calculated from $n_{2d}$ assuming a 3d Fermi surface, using $E_F = \hbar^2(3\pi^2 n_{2d}/t)^{2/3}/(2m^*)$ and $m^* \approx 0.15 m_e$ [35], and the ARPES $E_F$ is simply read off from the spectra. In Fig. 3(d), the first notable feature is the presence of a peak in the ARPES data around $x_T$ and its absence in the transport data. This can be understood by considering how the band structure changes near the critical point. Throughout the topological transition near $x_T$, as illustrated in Fig. 3(e), a dramatic change occurs around the Dirac point and the CB minimum, while at $E_F$ the change is only marginal with the overall size of the Fermi circle getting gradually reduced and the surface band smearing out. This explains why the transport $E_F$, which simply represents the size of the Fermi circle, does not show the kind of peak observed in the ARPES data. Another observation is that the ARPES $E_F$ is generally higher than the transport $E_F$. This is mainly due to the downward band bending on $Bi_2Se_3$ surfaces [35-36]. Therefore, the ARPES $E_F$, which measures the surface Fermi level, should always be higher than the transport $E_F$, which is an average value of the surface and bulk Fermi levels.

Finally, Figure 3(f) shows how the ARPES spectrum evolves with increasing x. For x < 3%, the surface band is clearly visible, implying that the system is a topological metal. Although a gap is formed at the Dirac point in the x = 4% spectrum, the surface bands seem to disappear completely only for x > 7% according to the transport analysis. The system becomes a 3d-VRH insulator for x > 15%, and a full band insulator for $x \gtrsim 25\%$. $(Bi_{1-x}In_x)_2Se_3$ is unique in that it is so far the only system that provides not only topological but also electronic tunabilities with a structural match to $Bi_2Se_3$ [37]. Therefore, it offers an ideal venue to fabricate tunable topological devices with seamless epitaxial tunnel barriers and gate dielectrics.

We would like to thank Leszek Wielunski and Thomas Emge for their technical assistance with RBS and XRD measurements respectively. Also, we are grateful for the discussions with Peter Armitage, Ronaldo Valdes Aguilar, and Sang-Wook Cheong. The work is supported by IAMDN of Rutgers

**Figure captions**

**Fig. 1. (color online) (a)** Typical RHEED images for x = 0, 20 and 95%. The $2^{nd}$-phase free vertical streaks indicate atomically flat films for the full range of x. **(b)** Lattice constant as a function of x obtained from the horizontal separation of the maxima in (a). **(c)** X-ray diffraction data for x = 0, 10, 80%. The (003n) peaks, marked by triangles for n = 1 – 6, provide a quintuple layer thickness of ~9.5 Å, almost independent of x.

**Fig. 2. (color online) (a)** Resistance vs. temperature for 60 QL films with various values of x. From bottom up x = 2.0, 8.0, 10, 12, 15, 16, 18, 20, and 25% **(b)** $R(1.5\ K)/R_{min}$ vs. x, exhibiting a metal-insulator transition at x ≈ 15%, defined as $x_I$. **(c)** and **(d)** show how resistance scales with temperature. Linearity in (c) and (d) implies, respectively, activation and 3d-VRH temperature dependence. **(e)** $T_0$ for the 3d-VRH (left axis) and $E_a$ for the activation (right axis) formula as a function of x. The vertical dotted line is placed at x = 25% as an approximate transition point ($x_B$) between VI and BI. **(f)** A photograph of 60 QL films with x = 0 and 80%.

**Fig. 3. (color online) (a)** Mobility (open squares/left axis) and carrier density (solid squares/right axis) vs. x at 1.5 K. Inset: $dR^N_{xy}/dB$ vs. B showing that the non-linearity disappears between x = 6 and 8%. **(b)** Mean free path, l, vs. x. As the mean free path becomes much smaller than half of the film thickness (dotted line), the effective dimensionality transforms to 3d. **(c)** $k_Fl$ vs. x. This value drops below one above x ≈ 15%. **(d)** The Fermi energy, $E_F$, relative to the CB minimum, vs. x, estimated from ARPES and transport. **(e)** A schematic of how the band structure changes near $x_T$. Top panel: Cross sections of the Fermi surface at $E_F$. Bottom panel: E vs. k band dispersion. The red solid (green dotted) lines represent the bulk conduction (surface) bands: for simplicity, the valence bands are not shown here. As x increases toward $x_T$, the bulk gap closes, and once x goes beyond $x_T$, the gap reopens and grows with increasing x; during this process, the conduction and valence bands invert and the surface band disappears. $x_T$ may not be a single point if the gap-opening at the Dirac point and disappearance of the surface band do not occur

simultaneously. **(f)** ARPES spectra for x = 2, 4, 8, 20, and 30%. TM, NM, VI, and BI represent topological metal, non-topological metal, variable range hopping insulator, and band insulator, respectively.

**Figure 1 (single column figure)**

(a)
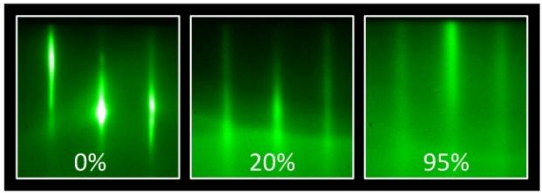

(b)
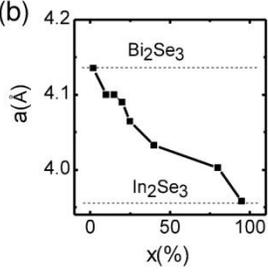

(c)
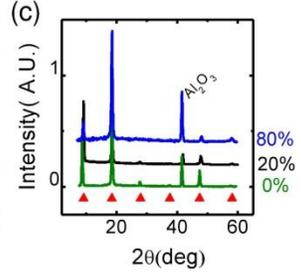

**Figure 2 (single column figure)**

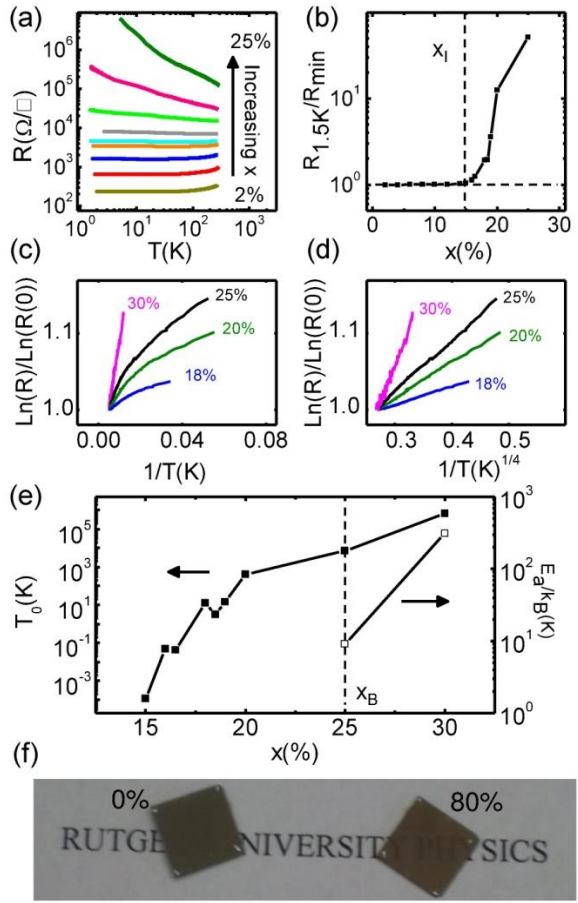

**Figure 3 (two column figure)**

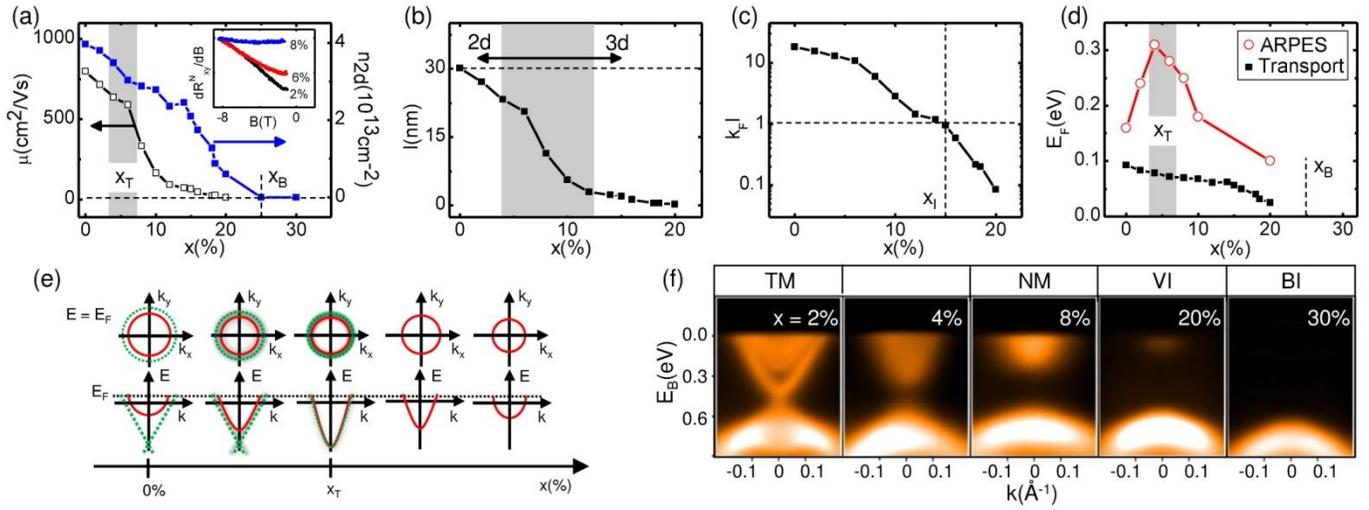